\documentclass[12pt,notitlepage]{article}
\usepackage{amsmath,amssymb}
\usepackage{graphicx}

\usepackage{color}

\def\ignore#1{{}}

\newcounter{sxn}

\newcounter{axn}

\newdimen\mybaselineskip
\newcommand{\beeq}{\begin{equation}}
\newcommand{\eneq}{\end{equation}}
\newcommand{\beqn}{\begin{eqnarray}}
\newcommand{\eeqn}{\end{eqnarray}}

\newcommand{\ba}{\begin{array}}
\newcommand{\ea}{\end{array}}

\newcommand{\tR}{\tilde{R}}

\newcommand{\be}{\begin{equation}}
\newcommand{\ee}{\end{equation}}
\newcommand{\bea}{\begin{eqnarray}}
\newcommand{\eea}{\end{eqnarray}}
\newcommand{\beal}{\setcounter{letter}{1} \begin{eqnarray}}
\newcommand{\eeal}{\addtocounter{equation}{1} \end{eqnarray}}
\newcommand{\none}{\nonumber \\}
\newcommand{\req}[1]{Eq.\ (\ref{#1})}

\newcommand{\larrow}{\,\,\,\,\hbox to 30pt{\rightarrowfill}
\,\,\,\,}
\newcommand{\slarrow}{\,\,\,\hbox to 20pt{\rightarrowfill}
\,\,\,}

\newcommand{\half}{{\frac{1}{2}}}

\def\t {\tilde}

\def\la{\raise.16ex\hbox{$\langle$}\lower.16ex\hbox{}  }
\def\ra{\, \raise.16ex\hbox{$\rangle$}\lower.16ex\hbox{} }

\def\psibar{ \psi \kern-.65em\raise.6em\hbox{$-$} \lower.6em\hbox{} }
\def\psibarb{ \psi \kern-.65em\raise.6em\hbox{$-$}  }

\def\t {\tilde}

\def\Ai {\hbox{Ai}}
\def\Bi {\hbox{Bi}}

\begin{document}

\thispagestyle{empty}





\begin{center}  
{\LARGE \bf   Quantum Mechanics of the Interior of the Russo-Susskind-Thorlacius Black Hole}

\vspace{1cm}

{\bf  Ramin G.~Daghigh$\sharp$, Michael D.\ Green$\star$, Gabor Kunstatter$\dagger$}
\end{center}

\centerline{\small \it $\sharp$ Natural Sciences Department, Metropolitan State University, Saint Paul, Minnesota, USA 55106}
\vskip 0 cm
\centerline{} 

\centerline{\small \it $\star$ Mathematics and Statistics Department, Metropolitan State University, Saint Paul, Minnesota, USA 55106}
\vskip 0 cm
\centerline{} 

\centerline{\small \it $\dagger$ Department of Physics, University of Winnipeg and}
\vskip 0 cm
\centerline{\small \it Winnipeg Institute for Theoretical Physics, Winnipeg, Manitoba, Canada R3B 2E9}
\vskip 0 cm
\centerline{} 

\vspace{1cm}
\begin{abstract}
We study the quantum mechanics of homogeneous black hole interiors in the RST model of 2D gravity. The model, which contains a dilaton and metric, includes radiation back-reaction terms and is exactly solvable classically. The reduced phase space is four dimensional. The equations for one pair of variables can be trivially solved.  The dynamics of the remaining degree of freedom, namely the dilaton, is more interesting and corresponds to that of a particle on the half line in a linear potential with time dependent coupling. We construct the self-adjoint extension of the corresponding quantized Hamiltonian and numerically solve the time dependent Schr$\ddot{\mbox{o}}$dinger equation for Gaussian initial data. As expected the singularity is resolved and the expectation value of the dilaton oscillates between a minimum and maximum, which both gradually decrease with time due to the time dependence in the potential.  In the classical black hole spacetime, the maximum value of the dilaton corresponds to the size of the horizon while the minimum is the singularity. The quantum dynamics, therefore, corresponds at the semi-classical level to an evaporating black hole. The rate of quantum fluctuations increases as the system evolves but intriguingly, at longer times the expectation value of the radius undergoes  ``revivals" in which the amplitude of oscillations between minimum and maximum temporarily increases. These revivals are also characteristic of the quantum dynamics of the  {\it time independent} quantum linear potential.

\end{abstract}

\newpage

\section{Introduction}
\vskip 1cm

Since quantum gravity is experimentally inaccessible and may remain that way for many years to come, one has to demand certain theoretically motivated criteria from any viable quantum gravity theory.  Resolving black hole singularities that appear in the classical theory is one of the most important of these criteria. In addition, a viable quantum gravity theory should be able to describe the endpoint of black hole evaporation and  resolve the information loss paradox.  In the early 1990's,  Callan et al.\cite{CGHS} proposed to study these issues in the context of  a two dimensional toy model subsequently dubbed the CGHS model. This model had the advantage that one can include Hawking radiation back-reaction in a relatively simple and rigorous form by computing the one-loop conformal anomaly for a set of quantized $N$ massless scalars. In the limit of large $N$, the one loop term is exact. Since then, a variety of classically solvable two-dimensional gravity models, including spherically symmetric gravity, have been used (for a review, see \cite{Grumiller}) to tackle basic questions of quantum gravity without having to deal with the technical complications that appear in the full higher dimensional theory.  The CGHS model didn't resolve the classical singularity and was not solvable once Hawking radiation was added.  In order to fix the latter problem,  Russo et al.\cite{RST} added a local term to the anomaly, but the singularity remained. In addition,  the model was shown to violate energy conservation in the form of an energy ``thunderbolt" that emanates from the endpoint of the collapse/radiation process.  This suggests, among other things, that the theory as given is not complete.  An important question is therefore  whether quantizing the gravitational degrees of freedom in the model can resolve the singularity and the other pathologies of the theory. 

In recent years, interest in the CGHS model was revived by the work of Ashtekar et al.\cite{Ashtekar1} who re-analyzed Hawking radiation in the model in the context of quantum geometry and argued that information was not lost. A more recent paper\cite{Ashtekar2}  did a numerical analysis of the semi-classical model that revealed interesting universal behavior not present in the purely classical case.

Of more direct relevance to the present work is the paper by Levanony and Ori\cite{Levanony} who analyzed the near singularity dynamics in the interior of a CGHS black hole quantum mechanically. They argued that the fields would tend to homogeneity in this limit, and showed that the resulting quantum theory resolved the singularity as required.
More recently Gegenberg et al.\cite{Kunstatter-RST} applied  an analysis similar to that of \cite{Levanony} to  the homogeneous interior of black holes in the RST model. In particular, they performed a complete analysis of the dynamics and the space of solutions, identifying the singularity and isolating the black hole sector. By first constructing the Hamiltonian for the reduced phase space dynamics, they were able to quantize the theory near the singularity and show that the singularity can indeed be resolved. This provided the first steps in a more complete quantization of this system.

The purpose of the present paper is to proceed further in this general program. We consider  dynamics of the dilaton field in the black hole interior using  an approximation in which the equation for the radiating degree of freedom is first solved classically. After a suitable canonical transformation, the remaining reduced Hamiltonian for the dilaton is equivalent to that of a particle on the half-line in a linear potential with time dependent, monotonically increasing, coupling\footnote{A pedagogical analysis  of the quantum dynamics of the time independent version of this ``bouncing-ball'' potential can be found in \cite{bouncing-ball}. }. We construct a self-adjoint Hamiltonian for the system, and solve the resulting time dependent Schr$\ddot{\mbox{o}}$dinger equation for Gaussian initial data that are meant to represent an initial semi-classical black hole. We verify the accuracy by using two different methods: Crank-Nicholsen and a spectral method. The calculations agree to numerical accuracy.

We note that the quantized linear potential on the half line is relevant to recent quantum measurements of neutrons in a gravitational potential. (See for example \cite{gravity-neutron}.) In addition, the time-dependent linear potential has numerous physical applications (see \cite{PChem-Feng} for more details).  To the best of our knowledge, the Schr$\ddot{\mbox{o}}$dinger equation on the half line with time dependent linear potential has not been considered previously. 
        
The paper is organized as follows: In the next Section we review the RST model and the analysis in \cite{Kunstatter-RST}. Section \ref{sec:DilatonQuantization} presents the Hamiltonian describing the dilaton dynamics and constructs the corresponding time dependent Schr$\ddot{\mbox{o}}$dinger equation. Section \ref{sec:numerics} describes the numerical calculation and exhibits the results.  Section \ref{sec:bounded} presents the numerical calculation in the case of a bounded coupling.  Finally, Section \ref{sec:conclusions} presents conclusions, speculations and prospects for future work.
\section{The Model}
\label{sec:model}

We consider  initially the classical CGHS model with $N$ conformally coupled massless scalar fields $f_i$, $i=1,2,...,N$:
\bea
I[g_{ij},\phi]&:=\frac{1}{2\pi}\int d^2x\sqrt{-g}e^{-2\phi}\left[R(g)+4(|\nabla\phi|^2+\lambda^2)\right]
   +\frac{1}{2\pi} \int d^2x \sqrt{-g}\sum_i |\nabla f_i|^2 ~.
\label{action0}
\eea
Quantizing the scalars yields the usual trace anomaly\cite{BirrelDavies}, which we add to the above action in a local form that was first introduced by Hayward\cite{hayward95}. We use the conventions and notations of \cite{Kunstatter-RST}. The local form of the action that forms the basis of our analysis is
\bea
I[g_{ij},\phi,z]&:=\frac{1}{2\pi}\int d^2x\sqrt{-g}\left\{e^{-2\phi}\left[R(g)+4(|\nabla\phi|^2+\lambda^2)\right]
\right.\none
&\left.+\frac{\kappa}{2}( R(g)z -\frac{|\nabla z|^2}{2}-R(g)\phi)\right\}~
\label{action1}
\eea
after setting the sources $f_i(x)$ to zero.  The first line in the above is the classical CGHS Lagrangian, with vacuum energy $\lambda^2$, whereas the second line represents the one-loop contribution from the conformal anomaly,
with $\kappa:=N/12$ ($\hbar$ has been set to one). In the limit of large $N$, the one loop contribution is exact. The last term in the second line is the local anomaly term added by RST\cite{RST} in order to make the semi-classical model solvable.

\subsection{Equations of Motion}
The equations of motion are given in Eqs.\,(3)-(5) of Hayward\cite{hayward95} with the sources $f_i(x)$ set to zero:
\bea
&A^- R_{\mu\nu}+2 A^+\nabla_\mu\nabla_\nu\phi
-\frac{\kappa e^{2\phi}}{4}\left(2\nabla_\mu\nabla_\nu z+\nabla_\mu z\nabla_\nu z-\half g_{\mu\nu}|\nabla z|^2\right)=0~;\label{met}\\
&A^+R+4\left(\nabla^2\phi-|\nabla\phi|^2+\lambda^2\right)=0~;\label{dil}\\
&\nabla^2z+R=0~,\label{z}
\eea
where
\be
A^\pm:=1\pm\frac{\kappa e^{2\phi}}{4}~.
\ee
One can formally recover the usual non-local form of the action by writing the solution to (\ref{z}) as
\be
z = -\frac{1}{\Box}R~.
\label{eq:z}
\ee
where $1/\Box$ ($\Box \equiv \nabla^2$) refers to the scalar Green's function. Substituting (\ref{eq:z})  back into the $z\Box  z$ term in the action (\ref{action1}) gives the usual non-local form $R\frac{1}{\Box}R$ of the Polyakov action. A more careful analysis\cite{hayward95} verifies that this heuristic process does indeed work.

The dynamical content of the theory can be understood as follows. There are initially five independent fields ($g_{\mu\nu}$, $z$, $\phi$). $z$ is effectively the radiation field and is zero in the absence of the radiation term in the action. There are two constraints associated with the diffeomorphism invariance that in turn are associated with two gauge degrees of freedom, which leaves a single propagating dynamical field theoretic degree of freedom. In the absence of the radiating field, the CGHS model has no propagating fields.

In the following we will be examining the homogeneous interior of a static black hole so that we need consider only quantum mechanics and not quantum field theory.  As will become apparent below, the physical phase space is four dimensional, consisting effectively of the black hole mass and its conjugate, as well as the black hole temperature and its conjugate. The mass and temperature are independent in this model.

We work in conformal gauge:
\be
ds^2 = e^{2\rho(t)}\left(-dt^2+dx^2\right) ~,
\ee
where $t$ and $x$ are spatial and time coordinates respectively.  In this gauge, the metric equations, \req{met}, reduce to\cite{Kunstatter-RST}:
\bea
-A^-\ddot\rho+2A^+(\ddot\phi-\dot\phi\dot\rho)+\frac{\kappa e^{2\phi}}{2}(-\ddot z+\dot z\dot\rho-\frac{1}{4}\dot z^2 )=0~;\label{mettt}\\
A^-\ddot\rho-2A^+\dot\phi\dot\rho+\frac{\kappa e^{2\phi}}{2}(\dot z\dot\rho-\frac{1}{4}\dot z^2 )=0~.\label{metxx}
\eea
The off-diagonal component of the Einstein equation is trivial in this case.
The dilaton equation of motion, \req{dil}, is
\be
A^+\ddot\rho-2\ddot\phi+2\dot\phi^2+2 \lambda^2 e^{2\rho}=0~.
\label{dilh}
\ee
Finally, the $z$ equation of motion, \req{z}, is simply 
\be 
\ddot z=2\ddot\rho~.
\label{eq:zddot}
\ee
By subtracting (\ref{mettt}) from (\ref{metxx}), and substituting (\ref{eq:zddot}), one obtains:
\be
\ddot \rho = \ddot \phi ~.
\label{eq:rhoddot}
\ee
Equations (\ref{eq:zddot}) and (\ref{eq:rhoddot}) are trivially solved to yield:
\bea
z(t)-2\rho(t) &=& z_1t+z_0~;
\label{eq:zSoln} \\
\rho(t) - \phi(t) &=& p_1t +p_0~,
\label{eq:rhoSoln}
\eea
which determine the radiating field and conformal mode of the metric in terms of four parameters $(z_1,z_0)$ and $(p_1,p_0)$.

Using (\ref{eq:zSoln}) and (\ref{eq:rhoSoln}), the dilaton equation (\ref{dilh}) and metric equation (\ref{metxx}) give the following two second order equations, respectively, for $\phi(t)$:
\bea
A^-\ddot{\phi} - 2\dot{\phi}^2 -2 p_1 A^- \dot{\phi} + \frac{\kappa e^{2\phi}}{2}\left(p_1^2-\frac{z_1^2}{4}\right)=0
~;\label{eq1}\\
-A^-\ddot\phi+2\dot\phi^2+2\lambda^2 e^{2(\phi+p_1 t +p_0)}=0\label{eq2}~.
\eea
 Clearly they cannot be independent. In fact there is a consistency condition that is essentially the Hamiltonian constraint, a consequence of time translation invariance.
Using the solutions (\ref{eq:zSoln}) and (\ref{eq:rhoSoln}) the consistency condition reduces to:
\be
-2p_1 A^-\dot\phi+e^{2\phi}\left[\frac{\kappa}{2}\left(p_1^2-\frac{z_1^2}{4}\right))+2\lambda^2 e^{2(p_1t+p_0)}\right]=0~.\label{consisteq}
\ee
As we will see, this constraint determines either $z_1$ or $p_1$ associated with (\ref{eq:zSoln}) or (\ref{eq:rhoSoln}), respectively, in terms of the constant of motion that results from integrating (\ref{consisteq}). Time translation invariance implies that either $p_0$ or $z_0$ can be set to zero without loss of generality, resulting in a  solution space that consists of four physical parameters.


\subsection{Classical Solutions}
First we make the field redefinition\cite{Kunstatter-RST},
\be
\t{R} := e^{-2\phi}~.
\ee
The dilaton field equation (\ref{eq2}) becomes
\be
\ddot{\tR} = -\frac{\kappa}{4}\frac{\dot{\tR}\dot{\tR}}{\tR(\tR-\kappa/4)}-4\lambda^2 e^{2(p_1t+p_0)}\frac{\tR}{\tR-\kappa/4}~.
\label{eq:Rddot}
\ee
It can easily be verified that  (\ref{eq:Rddot}) is generated by the following Hamiltonian:
\be
H_R=\frac{\Pi^2_{\tilde{R}}}{2}\left(\frac{\tilde{R}}{\tilde{R}-\kappa/4}\right)^2+4\lambda^2 e^{2(p_1 t+p_0)} \left(\tilde{R}-\frac{\kappa}{4}\ln \tilde{R} \right) ~.
\label{Hamiltonian1}
\ee 
Moreover, (\ref{eq:Rddot}) has the first integral
\bea
\frac{\dot \tR}{\tR} \left(\tR -\frac{\kappa}{4}\right) +\frac{2\lambda^2}{p_1} e^{2(p_1t+p_0)}=c_1 = \hbox{constant}.
\label{eq:FirstIntegral}
\eea
Comparing (\ref{eq:FirstIntegral}) to the consistency condition (\ref{consisteq}) requires $c_1 = p_1^2- \frac{z_1^2}{4}$. 

The general solution to \req{eq:Rddot} is\cite{Kunstatter-RST}
\be
\tR(t) = e^{-2\phi(t)}= e^{- W(f(t))+\frac{2}{\kappa}\theta(t)},
\label{phisol}
\ee
where $W(x)$ is the Lambert W function\cite{DLMF} defined implicitly by {$W(x)e^{W(x)}=x$} and
\bea
f(t)&:=&-\frac{4}{\kappa} e^{2\theta(t)/\kappa}~;\\
\theta(t)&:=&-2e^{-2\phi}-\kappa\phi\none
   &=&\frac{2\lambda^2}{p_1^2}e^{2(p_1 t+p_0)}+c_1 t+ \theta_0~.\label{gensol}
\eea
The Lambert W function has a branch point singularity at $x=-1/e$. This singularity corresponds to a curvature singularity in the metric, and occurs at $\tR = \kappa/4$.  On the principal branch of $W(x)$ for which $W(x)>W(-1/e)$, $W(x)\to \infty$ as $x\to\infty$. On the other branch $W(x)\to -\infty$ as $x\to 0$. The latter is the physical branch. The stationary black hole sector of the solution space corresponds to $c_1=0$. In this case there is a Killing horizon at finite $\theta = \theta_H$ where the metric component $e^{2\rho_H}=0$ and the curvature is finite. It was shown in \cite{Kunstatter-RST} that the solution can be analytically extended past this point, and a suitable radial coordinate $r$ defined in the exterior region such that $r\to \infty$ corresponds to the asymptotic exterior region of the black hole. It was shown that this solution corresponds to the RST solution\cite{RST} for $P=0$ and mass $M= \theta_0\lambda\sqrt{\kappa}$, which, as noted by Birnir and Giddings has a physical interpretation as a semi-classical black hole in thermal equilibrium with their environment at a fixed temperature $T_{BH}=\frac{p_1}{2\pi}$ that is independent of the mass. 

So far only the matter field that gave rise to the conformal anomaly has been quantized. We will now proceed to quantize the single reduced gravitational degree of freedom represented by the dilaton.  Note that we have solved for the radiation field and conformal mode of the metric classically so that, in particular, the parameter $p_1$, which corresponds to the black hole temperature $T_{BH}$ in the classical solution, does not fluctuate quantum mechanically. However, the dilaton and therefore the parameter $c_1$ will undergo fluctuations, so that the system will not be in equilibrium. The quantum dynamics will  therefore be considerably more interesting than the classical dynamics.

\section{Quantum Mechanics of the Dilaton Field}
\label{sec:DilatonQuantization}
We now proceed to quantize the dilaton field on the Hamiltonian constraint surface. 
It was shown in \cite{Kunstatter-RST} that one can quantize (\ref{Hamiltonian1}) in the limit $\tR\to\kappa$ and resolve the singularity to get a big bounce. 
In the following we will quantize (\ref{Hamiltonian1}) exactly.
We do this by first implementing a canonical transformation that significantly simplifies the Hamiltonian. We define:
\bea
y &=& \left(\tilde{R}-\frac{\kappa}{4} \ln \tilde{R}\right)-\eta ~;\nonumber  \\
\Pi_y &=& \Pi_{\tilde{R}}\left(\frac{\tilde{R}}{\tilde{R}-\kappa/4}\right)~,
\eea
where $\eta=\frac{\kappa}{4}-\frac{\kappa}{4} \ln \frac{\kappa}{4}$. The location of the black hole singularity at $\tilde{R}=\kappa/4$ corresponds to $y=0$, so that we will need to restrict the physical phase space to the half line $y>0$. In  terms of $y$ and its conjugate $\Pi_y$ the Hamiltonian becomes
\beeq
H_y=\frac{\Pi^2_y}{2}+4\lambda^2 e^{2(p_1t+p_0)} (y+\eta) ~.
\label{Hamiltonian2}
\eneq 
This resembles the Hamiltonian for a bouncing particle in a linear gravitational potential, with reflecting boundary conditions at $y=0$.  The key difference in our case is that the slope of the potential increases exponentially with time.  In the gravitational potential analog, this means the gravitational acceleration $g$ increases exponentially with time, which in turn results in the maximum height of the bounce decreasing with time. We will see that the corresponding expectation value in the quantum theory does precisely this, with interesting consequences for the quantum black hole. 

 The constant $\eta$ corresponds to a time dependent, but spatially independent, shift in the potential. We will see in the following that this can always be absorbed into a time dependent phase in the wave function that does not affect expectation values. It does, however, need to be taken into account when calculating the energy of the system as a function of time.

We quantize in the Schr$\ddot{\mbox{o}}$dinger representation, so that $\hat y = y$, with measure 
\be
\langle \psi_1 | \psi_2 \rangle = \int^\infty_0 dy \psi_1^*(y) \psi_2(y)~.
\ee
The boundary condition at $y=0$ implies that the conjugate to $\hat y$, namely $\hat{\Pi}_y= -i\hbar \partial_y$ does not exist as a self adjoint operator\cite{bonneau}.  A one parameter family of self-adjoint extensions of the Hamiltonian operator on $y\in [0,\infty]$  does exist, corresponding to the boundary conditions
\be
\psi(0) +L\psi'(0) =0 ~.
\ee 
For simplicity we set the the extension parameter $L=0$, i.e. choose Dirichlet boundary conditions, which are the natural boundary conditions to choose for the bouncing ball problem. In the case of quantum gravity, it is less obvious what the choice is, except that if one considers the infinite wall at $y=0$ to be a limiting case of a finite potential, it has been shown\cite{Robin} that Dirichlet boundary conditions are generic in the sense that obtaining any other boundary conditions as the infinite limit of a finite potential requires fine tuning of parameters as the limit is taken.

The task, then, is to solve the Schr\"{o}dinger equation of the form
\begin{eqnarray}
	i\frac{\partial \psi(y,t)}{\partial t}
	&=& H_y(t)\psi(y,t)\nonumber\\
	&=& \frac{1}{2}\left(-\frac{\partial^2}{\partial y^2} + V(y,t)\right)
	\psi(y,t)
	\label{eq:Shcrodinger1}
\end{eqnarray}
under Dirichlet boundary conditions for interesting initial data. This needs to be done numerically. The calculation is described in the next section.

\section{Numerical Calculation}
\label{sec:numerics}

We use two different numerical methods to solve this problem, the Crank-Nicholson method and a spectral method.  We first describe the details of the spectral method which follows the implementation in \cite{Sanjeev}.

 \subsection{Spectral Method}
We start by finding solutions to the instantaneous eigenvalue problem
\begin{eqnarray}
H_y(t)\psi_n(y,t) = E_n(t)\psi_n(y,t)
\label{eq:Schrodinger2}
\end{eqnarray}
with normalized eigenstates
\begin{eqnarray}
\int_0^\infty dy \psi^*_m(y,t)\psi_n(y,t) = \delta_{mn}~.
\end{eqnarray}
We then write
\begin{eqnarray}
\psi(y,t) = \sum_{n=0}^\infty c_n(t)e^{i\theta_n(t)}\psi_n(y,t)
\label{eq:expansion}
\end{eqnarray}
and choose {$\dot{\theta}_n(t) = - E_n(t)$}
so that (\ref{eq:Schrodinger2}) becomes
\begin{eqnarray}
\sum_{n=0}^\infty\left(\dot{c}_n e^{i\theta_n}\psi_n+c_n e^{i\theta_n}\dot{\psi}_n\right)=0~.
\end{eqnarray}
We now take the inner product by integrating the expression above with $\int_0^\infty dy \psi^*_m$ and use the orthonormality condition, which gives
\begin{eqnarray}
\dot{c}_me^{i\theta_m(t)} = -\sum_{n=0}^\infty  c_n(t) e^{i\theta_n(t)}\int_0^\infty dy \psi^*_m(y,t)\dot{\psi}_n(y,t)~.
\label{eq:cdot}
\end{eqnarray}
We then calculate the right hand side of (\ref{eq:cdot}) starting from
\begin{eqnarray}
\int_0^\infty dy \psi^*_m
\frac{d}{dt}\left(H_y\psi_n\right)&=&\int_0^\infty dy \psi^*_m\frac{d}{dt}\left(E_n\psi_n\right)
\end{eqnarray}
and using {$\dot{H}= \dot{V}/2$}, we get
\begin{eqnarray}
(E_m-E_n) \int_0^\infty dy \psi^*_m \dot{\psi}_n = \int_0^\infty dy \psi^*_m(y,t)\left(\dot{E}_n(t) - \frac{1}{2} \dot{V}(y,t)\right)\psi_n(y,t)~,
\label{eq:phidot}
\end{eqnarray}
where we have used the self-adjointness of the Hamiltonian:
\begin{eqnarray}
\int_0^\infty dy \psi^*_m H_y \dot{\psi}_n &=& \int_0^\infty dy (H_y\psi_m)^*\dot{\psi}_n\nonumber\\
  &=& E_m\int_0^\infty dy \psi^*_m\dot{\psi}_n ~.
\end{eqnarray}
 When $m=n$, \req{eq:phidot} becomes
\begin{eqnarray}
\int_0^\infty dy \psi^*_n(y,t)\left(\dot{E}_n(t) - \frac{1}{2} \dot{V}(y,t)\right)\psi_n(y,t)&=&0 ~,
\end{eqnarray}
which implies $\langle \dot{E}\rangle = \frac{1}{2} \langle \dot{V} \rangle$.
This is consistent with the basic relationship
\begin{eqnarray}
\frac{d}{dt}\langle E\rangle &=& \frac{d}{dt}\langle H_y\rangle
   \nonumber\\
 &=& \langle \frac{\partial H_y}{\partial t}\rangle
  + \frac{i}{\hbar}\frac{d}{dt}\langle [H_y,H_y]\rangle \nonumber\\
    &=& \langle \frac{\partial H_y}{\partial t}\rangle ~.
\end{eqnarray}
When $m\neq n$, one can use the orthonormality of the basis states to conclude that
\begin{eqnarray}
 \int_0^\infty dy \psi^*_m(y,t)\left(\dot{E}_n(t) \right)\psi_n(y,t)
 &=& \dot{E} \langle \psi_m | \psi_n \rangle \nonumber\\
 &=& 0 
\end{eqnarray}
so that \req{eq:phidot} gives
\begin{eqnarray}
\dot{c}_m(t) = \sum_{n\neq m} \chi_{mn}(t) c_n(t) ~,
\end{eqnarray}
where
\begin{eqnarray}
\chi_{mn}(t) &:=& \frac{1}{2} \frac{ e^{i(\theta_n(t)-\theta_m(t))}}{(E_m(t)-E_n(t))} \dot{V}_{mn}(t) ~,
\end{eqnarray}
$\dot{V}_{mn}$ is the matrix element
\begin{eqnarray}
\dot{V}_{mn}(t) &:=& \langle \psi_m|\dot{V}(t)|\psi_n\rangle \nonumber\\
   &=& \int_0^\infty dy \psi^*_m(y,t) \dot{V}(y,t) \psi_n(y,t)
\end{eqnarray}
and
\begin{eqnarray}
\theta_n(t) = -\int_0^t d\tilde{t}E_n(\tilde{t}) + \theta_n(0) ~. 
\end{eqnarray}
Note that $\theta_n(0)$ are arbitrary integration constants that do not affect the physical state since they introduce time and space independent phases that can always be absorbed into the basis functions $\psi_n(y,t)$ (cf. Eq.\ (\ref{eq:expansion})).

What we need to do now is solve for the $\psi_n(y,t)$, $E_n(t)$, and $V_{mn}(t)$ for our model. 
In our case, we wish to solve the problem with a linear potential,
\begin{eqnarray}
i\frac{\partial \psi(y,t)}{\partial t}
    &=& \frac{1}{2} \left(-\frac{\partial^2}{\partial y^2} + f(t) (y+\eta) \right)
      \psi(y,t) ~.
\label{eq:linearSchrodinger}
\end{eqnarray}
The Eigenvalue problem at fixed $t$ then becomes
\begin{eqnarray}
\frac{1}{2}\left(-\frac{\partial^2}{\partial y^2} + f(t) y \right) \psi_n(y,t)
 = \left(E_n(t)-\frac{1}{2}f(t)\eta\right) \psi_n(t) ~.
\end{eqnarray}
We can see that the term involving $\eta$ just contributes a time dependent shift to the linear potential and can be absorbed into the energy term by redefining $E_n(t)$.  This introduces a time dependent phase change to $\psi$ that does not affect expectation values.  Since we are interested in the expected value of position, we take $\eta=0$ without loss of generality.  Note, however, the contribution from the $\eta$ term needs to be included when calculating the total energy as a function of time. 
The above expression can be modified by defining $x=f^{1/3}y$:
\begin{eqnarray}
\left(-\frac{\partial^2}{\partial x^2} + x \right) \psi_n(x) 
  = \lambda_n \psi_n(x) ~.
\label{eq:eigenEqn}
\end{eqnarray}
Note that the time dependence has disappeared from the Eigenvalue equation, i.e. it has been absorbed into the coordinate $x$, so that the eigenfunctions $\psi_n(x)$ and eigenvalues $\lambda_n=2f^{-2/3}(t)E_n(t)$ are independent of time.

\begin{figure}[tb]
	\begin{center}
	\includegraphics[height=5.3cm]{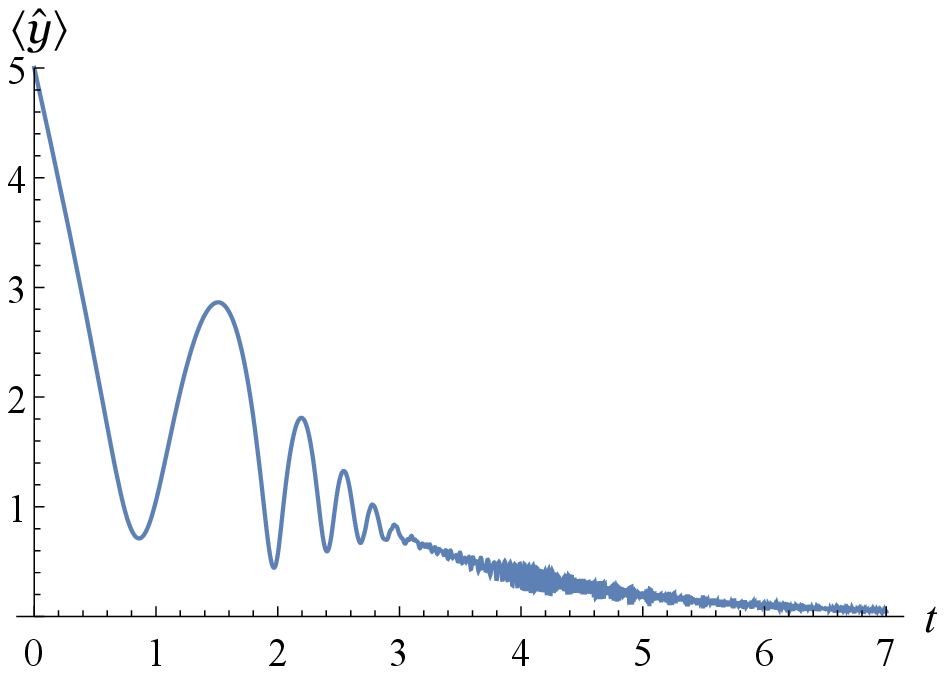}\qquad
	\includegraphics[height=5.3cm]{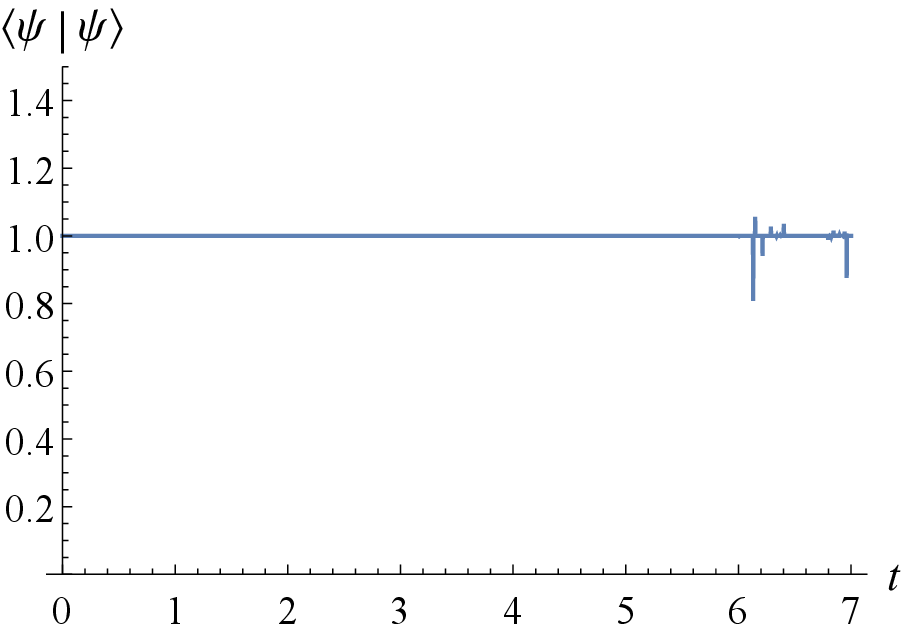}
	\end{center}
	\caption{Expectation value of the position and the norm of the Gaussian wavepacket are plotted as a function of time $t$.  We take $\eta=0$ and $2\lambda e^{p_0}=p_1=1$.  The units we use are $\hbar=c=1$.}
	\label{exgraph}
\end{figure}

The general solution to (\ref{eq:eigenEqn}) is
\begin{eqnarray}
\psi_n(x)&=& B_1\hbox{Ai}(x-\lambda_n) + B_2 \Bi(x-\lambda_n) ~,
\end{eqnarray}
where the $\Ai(x)$ and $\Bi(x)$ are Airy functions of the first and second kind. Since $\Bi(x)$ diverges as $x\to \infty$, the requirement of normalizability implies that $B_2=0$.
The eigenfunctions are, therefore,
\begin{eqnarray}
\psi_n(y,t)&=& B_n(t)\Ai(f^{1/3}y - \lambda_n) ~,
\end{eqnarray}
where the normalization factors are determined from 
\begin{eqnarray}
1&=&\left|B_n(t)\right|^2 \int_0^\infty dy\left|\Ai(f^{1/3}y - \lambda_n)\right|^2\nonumber\\ 
&=& \left|B_n(t)\right|^2f^{-1/3}(t) h_n^2 ~.
\end{eqnarray}
Here,
\begin{eqnarray}
 h_n^2&=& \int_0^\infty dx\left|\Ai(x - \lambda_n)\right|^2
\end{eqnarray}
is a time independent number.

The eigenvalues $\lambda_n$ are determined from the boundary conditions needed to make the operator $\partial^2/\partial x^2$ self-adjoint on the half-line. The general Robin boundary conditions are
\begin{eqnarray}
\psi_n(0,t) + L \left.\frac{\partial \psi_n(y,t)}{\partial y}\right|_{y=0} =0 ~.
\end{eqnarray}
Since we know explicitly the time dependence of $E_n(t)$, i.e.
\begin{eqnarray}
E_n(t) = \frac{1}{2}f^{2/3}(t)\lambda_n ~,
\end{eqnarray}
we can calculate
\begin{eqnarray}
\theta_n(t) = -\lambda_n \int_0^t dt f^{2/3}(t) ~.
\end{eqnarray}
We can also calculate $V_{mn}$, since
\begin{eqnarray}
\dot{V}_{mn}(t)
   &=& \dot{f}B_m(t)B_n(t)\int_0^\infty dy \Ai(f^{1/3}y- \lambda_m)y \Ai(f^{1/3}y-\lambda_n)\nonumber\\
  &=& \frac{\dot{f}f^{-1/3}}{h_nh_m} \int_0^\infty dx \Ai(x-\lambda_m)x\Ai(x-\lambda_n) ~.
\end{eqnarray}

In the case $f(t)=2e^{2t}$, in which we have taken $\eta = 0 $ and $2\lambda e^{p_0} = p_1 = 1$, we can solve for $\psi$ by starting with a series combination of Airy functions, $\psi_n$, that approximates the Gaussian wavepacket given by
\beeq
\psi(y,t=0)=\frac{\sqrt{2}}{\pi^{1/4}\sqrt{1 + \rm{erf}(5)}}\left[e^{-\frac{(y-5)^2}{2}}\right] e^{-5iy} ~,
\label{initialGaussian}
\eneq  
where ${\rm{erf}}(x) = \frac{2}{\sqrt{\pi}} \int_0^x e^{-t^2} \, dt$
is the error function.

The expected value of $\hat{y}$ is shown in Fig.\ {\ref{exgraph}.  The factor $e^{-5iy}$ in Eq.~(\ref{initialGaussian}) gives the wavepacket an initial velocity toward the singularity, which allows the bouncing behavior in $\left<\hat{y}\right>$ to happen earlier. Due to the high number of oscillations in the solution, we also verified that the norm of the wavefunction remained constant as a check that our numerical method was behaving properly.  The norm is also shown in Fig.~{\ref{exgraph}}.

\begin{figure}[tb]
	\begin{center}
	\includegraphics[height=5.3cm]{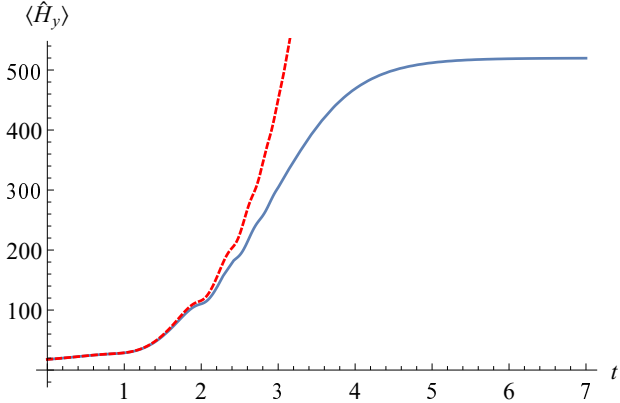}
	\end{center}
	\caption{The expectation value of the Hamiltonian versus time $t$.  The Red dashed line represents our exponentially increasing linear potential.  The solid line represents the case where the coupling is bounded.  We take $\eta=0$, $2\lambda e^{p_0}=p_1=1$ and $\Lambda = \sqrt{500} \lambda$.  The units we use are $\hbar=c=1$.}
	\label{exenergygraph-limited}
\end{figure}
\begin{figure}[tb]
	\begin{center}
	\includegraphics[height=5.3cm]{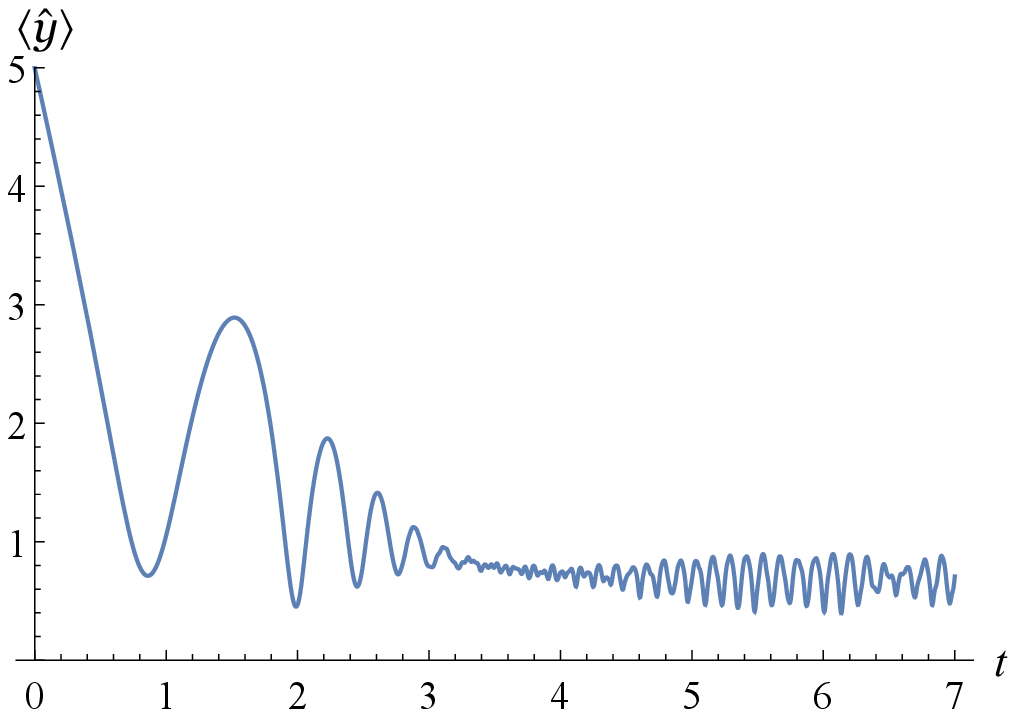}\qquad
	\includegraphics[height=5.3cm]{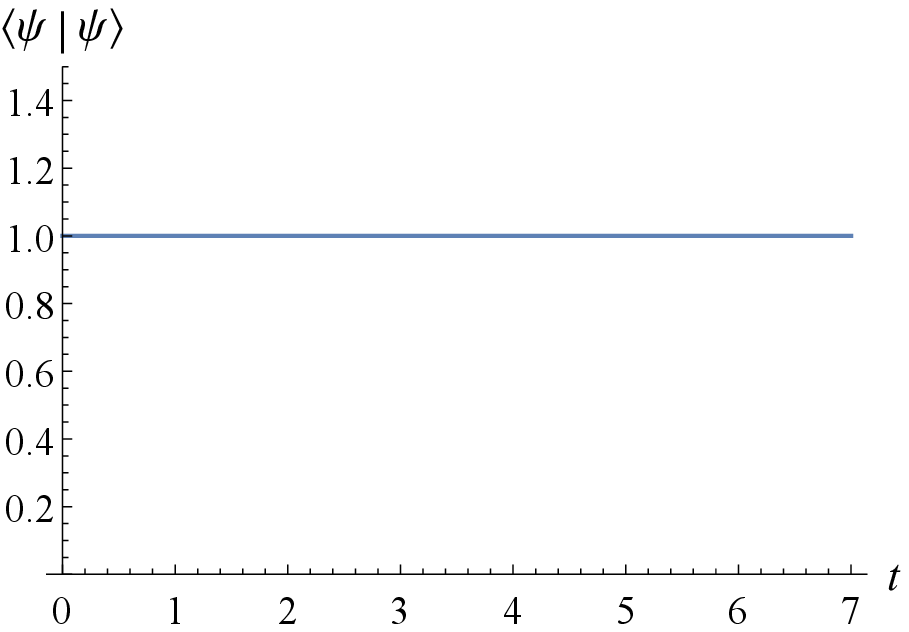}
	\end{center}
	\caption{Expectation value of the position and the norm of the wavepacket are plotted as a function of time $t$ for bounded coupling.  We take $\eta=0$, $2\lambda e^{p_0}=p_1=1$ and $\Lambda = \sqrt{500} \lambda$.  The units we use are $\hbar=c=1$.}
	\label{exgraph-limited}
\end{figure}

\subsection{Crank-Nicholson Method}
The Crank-Nicholson method is a well-known finite difference method used to find numerical solutions to partial differential equations.  In this case, to apply the boundary condition of an infinite wall at $y=0$ we replace the time dependent linear potential $4\lambda^2 e^{2(p_1t+p_0)} (y+\eta)$ with $4\lambda^2 e^{2(p_1t+p_0)} (|y|+\eta)$.  We now can allow an initial wave of the form
\beeq
\psi(y,t=0)=\frac{1}{\pi^{1/4}\sqrt{1 - e^{-25}}}\left[e^{-\frac{(y-5)^2}{2}} e^{-5iy}- e^{-\frac{(y+5)^2}{2}}e^{5iy}\right]~
\label{initialGaussianCN}
\eneq
to approach the singularity at $y=0$.  The above wavepacket is composed of two anti-symmetric Gaussian wavepackets, which are located on the positive and negative $y$-axis equi-distant from $y=0$.  Again, we have included factors of $e^{\pm 5iy}$ to give the wavepackets an initial velocity toward the singularity.   
As the two Gaussian wavepackets interact, the result on the interval $[0, \infty)$ is identical to the behavior of a bouncing Gaussian wavepacket on an infinite potential at $0$. 
We found that the solutions using the Crank-Nicholson method converged to those of the spectral method as we decreased the step size.


\section{Bounded Coupling}
\label{sec:bounded}

The coupling in the linear potential in Eq.~(\ref{Hamiltonian2}) diverges exponentially as time evolves.  This divergence appears due to the fact that in deriving the Hamiltonian (\ref{Hamiltonian1}) we treated the field $\rho(t)-\phi(t)$ in Eq.~(\ref{eq:rhoSoln}) classically.  
In a fully quantized quantum model,  such divergences should not appear.  
  Such a quantization is beyond the scope of the present work,  but in order to see qualitatively what might happen in the absence of such a divergence, we replace the term $\lambda^2 e^{2p_1 t}$ with the regularized form:
\beeq
\frac{\lambda^2 e^{2p_1 t}}{1+\frac{\lambda^2 e^{2p_1 t}}{\Lambda^2}} ~.
\eneq  
In Fig.~\ref{exenergygraph-limited}, we plot the expected value of the Hamiltonian (\ref{Hamiltonian2}) for the unbounded and bounded cases.  In Fig.~\ref{exgraph-limited}, we plot the expected position of the wavepacket of the dilaton field along with the norm of the wavepacket. Note that in the bounded case, the expected position of the wavepacket oscillates but no longer decreases to zero. This suggests that under this scenario the end point of the radiation would be a stable remnant rather than complete evaporation as occurs in the unregulated case.

We have verified that the precise form of the regularized coupling term does not qualitatively change the above picture, although the details of the transition to the steady state do change somewhat.

\section{Summary and Conclusion}
\label{sec:conclusions}

We have shown that the quantum dynamical evolution of the dilaton field in the interior of a homogeneous RST black hole is determined by a Schr\"{o}dinger equation with a linear potential with time dependent coupling on the half line.  We used two different numerical methods, spectral and Crank-Nicholson, to determine the evolution of a Gaussian wavepacket of the dilaton field. The expectation value of the wavepacket resembles that of a bouncing ball in the presence of an increasing gravitational field.  The maximum height of the bounce, which classically determines the horizon radius for the black hole, decreases with time.  The quantum dynamics, therefore, appear to provide an intriguing description of an evaporating black hole.  We also observe in Fig.~\ref{exgraph} that as the expectation value decreases, the oscillation frequency increases.    At late times in the process, the expectation value of the radius undergoes  ``revivals" in which the amplitude of oscillations between minimum and maximum temporarily increases, but within the numerical accuracy of the calculation, the maximum appears to decrease smoothly to zero, suggesting that the black hole evaporates completely.    We have also looked at an alternative scenario where the time dependent coupling is bounded.  This leads to a semi-classical description of black hole evaporation in which a stable finite radius is approached asymptotically. In this case the frequency of the oscillations approaches a constant and the interesting behavior in the amplitude of the oscillations, corresponding  to pulsations of the black hole radius, is more pronounced and easier to resolve numerically.  See Fig.~\ref{exgraph-limited}.  A similar pattern of revivals, or pulsations, as those mentioned above was noticed earlier in the context of a time independent linear potential in \cite{bouncing-ball}.  
In fact, the late time behavior observed in Fig.~\ref{exgraph-limited} is very similar to that seen in \cite{bouncing-ball}, which is not surprising since for late times the bounded coupling is nearly time independent.

We note that the exponential increase in energy that we have observed at long times is a direct result of the exponential  growth of the linear potential in the  Hamiltonian. This energy increase does not have direct physical significance for several reasons. First, the Hamiltonian in Eq.\ (\ref{Hamiltonian2}) was not derived directly from a phase space reduction of the full model. It was instead constructed to yield the correct dynamics for the dilaton. One is therefore free to add to it an arbitrary function of time that can be used to cancel this long term exponential growth.
Second, one of the peculiarities of  the original RST model is the behavior of the energy. For generic values of the solution parameters, the energy of the radiation field extends to infinity so that the ADM energy is not finite\cite{RST}. 
Moreover,  the semi-classical formation and evaporation of an RST black holes results in a naked singularity and potential emission of a ``thunderbolt" of infinite energy.  This rather unphysical property of the model was the main motivation for abandoning the model in the 1990's, but as noted in \cite{RST}, it may be cured by a full quantum treatment. 

As mentioned previously, we have quantized only the dilaton $\phi$, treating the dynamical degree of freedom associated with the radiation field classically. It is clearly of interest to do a more complete quantization of both the dilaton and the radiation field. Such an investigation, which is considerably more challenging and may require novel techniques to obtain results, is currently under investigation.

\vskip .5cm

\leftline{\bf Acknowledgments}
Gabor Kunstatter gratefully acknowledges the support of the Natural Sciences and Engineering Research Council of Canada. The authors would also like to thank Jack Gegenberg for useful discussions.



\def\jnl#1#2#3#4{{#1}{\bf #2} #3 (#4)}

\def\Zphys{{\em Z.\ Phys.} }
\def\jssc{{\em J.\ Solid State Chem.\ }}
\def\jpsJ{{\em J.\ Phys.\ Soc.\ Japan }}
\def\ptps{{\em Prog.\ Theoret.\ Phys.\ Suppl.\ }}
\def\PTP{{\em Prog.\ Theoret.\ Phys.\  }}
\def\LNC{{\em Lett.\ Nuovo.\ Cim.\  }}

\def\JMP{{\em J. Math.\ Phys.} }
\def\NPB{{\em Nucl.\ Phys.} B}
\def\NP{{\em Nucl.\ Phys.} }
\def\PLB{{\em Phys.\ Lett.} B}
\def\PL{{\em Phys.\ Lett.} }
\def\PRL{\em Phys.\ Rev.\ Lett. }
\def\PRA{{\em Phys.\ Rev.} A}
\def\PRB{{\em Phys.\ Rev.} B}
\def\PRD{{\em Phys.\ Rev.} D}
\def\PR{{\em Phys.\ Rev.} }
\def\PRe{{\em Phys.\ Rep.} }
\def\AP{{\em Ann.\ Phys.\ (N.Y.)} }
\def\RMP{{\em Rev.\ Mod.\ Phys.} }
\def\ZPC{{\em Z.\ Phys.} C}
\def\SCI{\em Science}
\def\CMP{\em Comm.\ Math.\ Phys. }
\def\MPLA{{\em Mod.\ Phys.\ Lett.} A}
\def\IJMPA{{\em Int.\ J.\ Mod.\ Phys.} A}
\def\IJMPB{{\em Int.\ J.\ Mod.\ Phys.} B}
\def\cmp{{\em Com.\ Math.\ Phys.}}
\def\JPA{{\em J.\  Phys.} A}
\def\CQG{\em Class.\ Quant.\ Grav.~}
\def\ATMP{\em Adv.\ Theoret.\ Math.\ Phys.~}
\def\AJP{\em Am.\ J.\ Phys.~}
\def\PRSA{{\em Proc.\ Roy.\ Soc.} A }
\def\ibid{{\em ibid.} }
\vskip 1cm

\leftline{\bf References}

\renewenvironment{thebibliography}[1]
        {\begin{list}{[$\,$\arabic{enumi}$\,$]}  
        {\usecounter{enumi}\setlength{\parsep}{0pt}
         \setlength{\itemsep}{0pt}  \renewcommand{\baselinestretch}{1.2}
         \settowidth
        {\labelwidth}{#1 ~ ~}\sloppy}}{\end{list}}


\end{document}